\documentclass[10pt,article]{IEEEtran}
\usepackage{graphicx}
\usepackage{algorithm2e}
\usepackage{multirow}
\newcommand{\mr}[1]{\multirow{2}*{#1}}

\newcommand{\RelaxFloats}{
       \renewcommand{\topfraction}{0.9}
       \renewcommand{\floatpagefraction}{0.9}
       \renewcommand{\textfraction}{0.1}
}

\begin{document}

\RelaxFloats


\title{CSI: A Paradigm for Behavior-oriented Delivery Services in Mobile Human Networks}

\author{\authorblockN{Wei-jen Hsu$^{1}$, Debojyoti Dutta$^{2}$, and Ahmed Helmy$^{1}$}
\authorblockA{\\$^{1}$Department of Computer and Information Science and Engineering, University of Florida $^{2}$Cisco Systems, Inc.
\\Email: $^{1}$ \{wjhsu, helmy\}@ufl.edu, $^{2}$dedutta@cisco.com}
}

\maketitle

\begin{abstract}

We propose behavior-oriented services as a new paradigm of
communication in mobile human networks. Our study is motivated by the
tight user-network coupling in future mobile societies. In such a
paradigm, messages are sent to inferred behavioral profiles, instead of
explicit IDs. Our paper provides a systematic framework in providing
such services. First, user behavioral profiles are constructed based on
traces collected from two large wireless networks, and their spatio-temporal
stability is analyzed. The implicit relationship discovered between mobile users
could be utilized to provide a service for message delivery and
discovery in various network environments. As an example application,
we provide a detailed design of such a service in
challenged opportunistic network architecture, named CSI. We provide a fully
distributed solution using {\it behavioral profile space} gradients and small world
structures.

Our analysis shows that user {\it behavioral profiles} are
surprisingly stable, i.e., the similarity of the behavioral profile of a user to its
future behavioral profile
is above 0.8 for two days and 0.75 for one week, and remains above 0.6 for five weeks.
The {\it correlation coefficient} of the similarity metrics 
between a user pair at different time instants is above 0.7 for four days,
0.62 for a week, and remains above 0.5 for two weeks.
Leveraging such a stability in user behaviors,
the CSI service achieves delivery rate very close to the delay-optimal 
strategy (above $94\%$), with minimal overhead (less than $84\%$ of the optimal).
We believe that this new paradigm will act as an enabler of multiple new services in mobile
societies, and is potentially applicable in server-based, heterogeneous
or infrastructure-less wireless environments.
\end{abstract}

\section{Introduction}  \label{sec:intro}

We envision future networks that consist of numerous ultra portable devices delivering highly personalized, context-aware services to mobile users and societies. Such scenarios elicit strong, tight-coupling between user behavior and the network. Users' mobility and on-line activities significantly impact wireless link characteristics and network performance, and at the same time, the network performance can potentially influence user activities and behavior. Such a tight user-network coupling provides a rich set of opportunities and poses several challenges. On one hand, fundamental understanding of the mobile user behavior becomes crucial to the design and analysis of future mobile networks. On the other hand, novel services can now be introduced and utilize such a coupling to effectively navigate mobile societies, providing efficient information dissemination, search and resource discovery.

In this paper, we propose a novel behavior-driven communication paradigm to enable a new class of services in mobile societies. Current communication paradigms, including unicast and multicast, require explicit identification of destination nodes (through node IDs or group membership protocols), while directory services {\em map} logical, interest-specific queries into destination IDs where parties are then connected using interest-oblivious protocols. The power and scalability of such conventional paradigms might be quite limited in the context of future, highly dynamic mobile human networks, where it is desirable in many scenarios to support implicit membership based on interest. In such scenarios, membership in interest-groups is not explicitly expressed by users, it is rather implicitly and autonomously inferred by network protocols based on behavioral profiles. This removes the dependence on third parties (e.g. directory lookup), maintenance of group membership (e.g., in multicast) or the need to flood user interests to the whole network, and minimizes delivery overhead to uninterested users.
 
Applying such a behavior-driven paradigm in mobile networks poses several research challenges. First, how can user behavior be captured and represented adequately? Second, is user behavior stable enough to enable meaningful prediction of future behavior with a
short history? How can such services be provided when the interest or behavior cannot
be centrally monitored and processed? And finally, can we design
privacy-preserving services in this context?

To address these questions we propose a systematic framework with two phases 1) behavioral profile extraction by analyzing large-scale
empirical data sets, investigating the stability of users in the behavioral space, and 2) leverage the behavioral profiles for service design -- We use the implicit structure in the human networks to guide message and query dissemination given a target profile.

Specifically, we first analyze network activity traces and design a summary of
user {\it behavioral profiles} based on the {\it mobility preferences}. 
The similarity of the {\it behavioral profile} for a given user to its future profile is high,
above $0.75$ for eight days and remains above $0.6$ for five weeks.
The surprising observation is that, the similarity metric between a pair of
users predicts their future similarity reasonably well. The correlation coefficient between
their current and future similarity metrics is above $0.7$ for four days, and remains 
above $0.5$ for fifteen days.

This phenomenon demonstrates that the {\it behavioral profile} we design is an intrinsic
property of a given user and a valid representation of the
user for a good period of time into the future. 
We refer to this phenomenon as the {\it stability} 
of user {\it behavioral profiles}, which
can be used to map the users into a high dimensional {\it behavioral space}.
The {\it behavioral space} is defined as a space where each 
dimension reflects a particular interest. For example, when 
we consider mobility preferences, each dimension represents 
the fraction of time spent at a given location. The position 
of users in the behavioral space reflects how similar they 
are with respect to the behavioral profile we construct.
We propose a new
communication paradigm, in which a {\it target profile} is used to replace
network IDs to indicate the intended receiver(s) of a message (i.e., those with 
{\it matching} behavioral profile to the target profile chosen by the sender are
the intended receivers.).
It is a {\it C}ommunication paradigm in human networks based on
the {\it S}tability of the user behavioral profile to discover the 
receivers {\it I}mplicitly, abbreviated as {\it CSI}. 
We present two modes of operation under the over-arching paradigm:
the {\it target mode (CSI:T) } and the {\it dissemination mode (CSI:D)}. The {\it target mode}
is used when the {\it target profile} is specified in the 
same context as the {\it behavioral profile} (i.e., the {\it target profile} is in terms
of {\it mobility preferences}). The {\it dissemination mode},
on the other hand, is used when the {\it target profile} is de-coupled from mobility preferences.

We show that our CSI schemes perform
very close to the delay-optimal schemes assuming global knowledge and improve
significantly over the baseline dissemination schemes. For the
{\it CSI:T mode}, comparing with the delay-optimal protocol,
our protocol is close in terms of success rate (more than $94\%$)
and has less overhead (less than $84\%$ to the optimal), and the delay is about $40\%$ more.
For the {\it CSI:D mode}, our protocol features lower
storage overhead than the delay-optimal protocol with more than $98\%$
success rate -- {\it
CSI:D} uses a storage overhead less than $60\%$ of the delay-optimal
protocol, while the delay of {\it CSI:D} is about $32\%$ more than the optimal.

\noindent{\textbf{\underline{Our Contributions}}}

\noindent (1) We introduce the notion of multi-dimensional {\it behavioral space}, and
devise a representation of user {\it behavioral profiles} to map users into
the behavioral space. Our study is the first to establish conditions for stability
of the relationship between campus users in this space.

\noindent (2) We propose {\it CSI}, a new communication paradigm
delivering message based on user profiles. The target profile in CSI can even be 
independent of the context of behavioral profile we use to construct the 
{\it behavioral space}.

\noindent (3) We design an efficient dissemination protocol utilizing the stability
of behavioral profiles and SmallWorld in mobile societies, then
empirically evaluate and validate the efficacy of our proposal using
large-scale traces from university campuses.

The outline of the rest of the paper is as follows.
We discuss the related work in section~\ref{sec:related} and important
background in section~\ref{sec:back}. This is followed by an analysis
to understand the user behavioral pattern in
section~\ref{sec:mobipattern}. We further discuss the potential usages
of this understanding in section~\ref{sec:transition} and
design our {\it CSI} schemes in section~\ref{sec:proto} as an example.
We use simulations to evaluate the performance of CSI schemes in
section~\ref{sec:exp}. Finally, we discuss some finer points 
in section~\ref{sec:disc} and conclude in section~\ref{sec:conc}.

\section{Related Work} \label{sec:related}

We conduct the first detailed systematic study on the spatio-temporal stability of user behaviors in mobile societies, a new dimension that has not been considered before. We lay the foundation of this work on a solid analysis of empirical user behaviors, enabled by extensive collections of user behavioral traces. Many of them can be found in the archives at \cite{MobiLib-web, CRAWDAD-web}. Our effort on the extraction of behavioral profiles and behavior-based user classification is related to the reality mining project~\cite{reality-mining} and the work by Hsu et al.~\cite{MOBI07} and Ghosh et al.~\cite{BU-classify}. We leverage the representation of mobility preference matrix defined by Hsu et al.~\cite{MOBI07}, which reveals more detailed user behavior than the five categories representation used in the reality mining~\cite{reality-mining} and the presence/absence encoding vector used by Ghosh et al.~\cite{BU-classify}.

In centralized trace analysis, the capability of classifying users based on their mobility preferences~\cite{MOBI07} or periodicity~\cite{classify-Dart} could potentially lead to applications such as behavior-aware advertisements or better network management. While understanding user behavior for these applications has its own merit, applications in centralized scenario (where user behaviors are collected, processed and mined at an aggregation point) are not our major focus in the paper.

The major application considered in this paper is to design a message dissemination scheme in decentralized environments. While several previous works exist in the delay tolerant network field, most of them (e.g. \cite{M-space-routing, epidemic, PRoPHET, socialnet-routing, group-routing}) consider one-to-one communication pattern based on network identities. The one-to-many communication targeted at a behavioral group presented in this paper is a new paradigm in decentralized environments. Some of the previous work assume existing infrastructure: PeopleNet~\cite{peoplenet} uses specialized geographic zones for queries to meet. The queries are delivered to randomly chosen nodes in the corresponding zone through the infrastructure. Others (e.g., \cite{PRoPHET, group-routing}) rely on persistent control message exchanges (e.g., the delivery probability) for each node to learn the structure of the network, even when there is no on-going traffic. From the design point of view, our approach differs from them by avoiding such persistent control message exchanges to achieve better power efficiency, an important requirement in decentralized networks.  


The spirit of our design is more similar to the work by Daly et al.~\cite{socialnet-routing}, in which each node learns the structure of the network locally and uses the information for message forwarding decisions. They use the SmallWorld network structure~\cite{smallworld} which often exists in human networks (as has been investigated in \cite{group-study, social-interaction}) and push the message toward nodes with high centrality to improve the chance of delivery. However, the learning process still involves message exchanges about past encounters, even in the absence of actual traffic. Our work, on the other hand, relies on the intrinsic behavioral pattern of individual nodes to ``position'' themselves in the behavioral space in a localized and fully distributed manner, without exchanging encounter history between nodes. The use of user behavioral profiles to understand the structure of the space is similar to the mobility space routing by Leguay et al.~\cite{M-space-routing} and the utility-based routing by Aiklas et al.~\cite{socialcast}. The major differences between this work and \cite{M-space-routing, socialcast} are two fold: First, we design the CSI:D mode, in which the target profile need not be related to the behavioral profile based on which the message dissemination decisions are made. Second, we also provide a non-revealing option in our protocol, thus no node has to explicitly reveal its behavioral pattern or interests to others, as opposed to~\cite{M-space-routing, socialcast}. The idea of merging similar users into a group based on their behavior has also been proposed in a two-tiered routing structure~\cite{group-routing}.


Another related paper is the work by Hsu et al.~\cite{Pcast} where the
authors focus on only sending messages to users with similar
behavioral profile to the sender. In this paper we introduce the notion of 
the {\it target profile} to decouple the behavioral profile of the sender from the 
destination profile in the message . This significantly
enhances the capability of the message dissemination schemes, by
allowing the sender to specify target behavioral profile (in CSI:T mode), or even
some target profiles that are orthogonal to the behavior
based on which we measure the similarity between users (in CSI:D mode).

\section{Background} \label{sec:back}

\subsection{Mobility-based User Behavior Representation} \label{sec: classify}

\begin{figure}
\centering
\includegraphics[width=2.5in, height=1.2in]{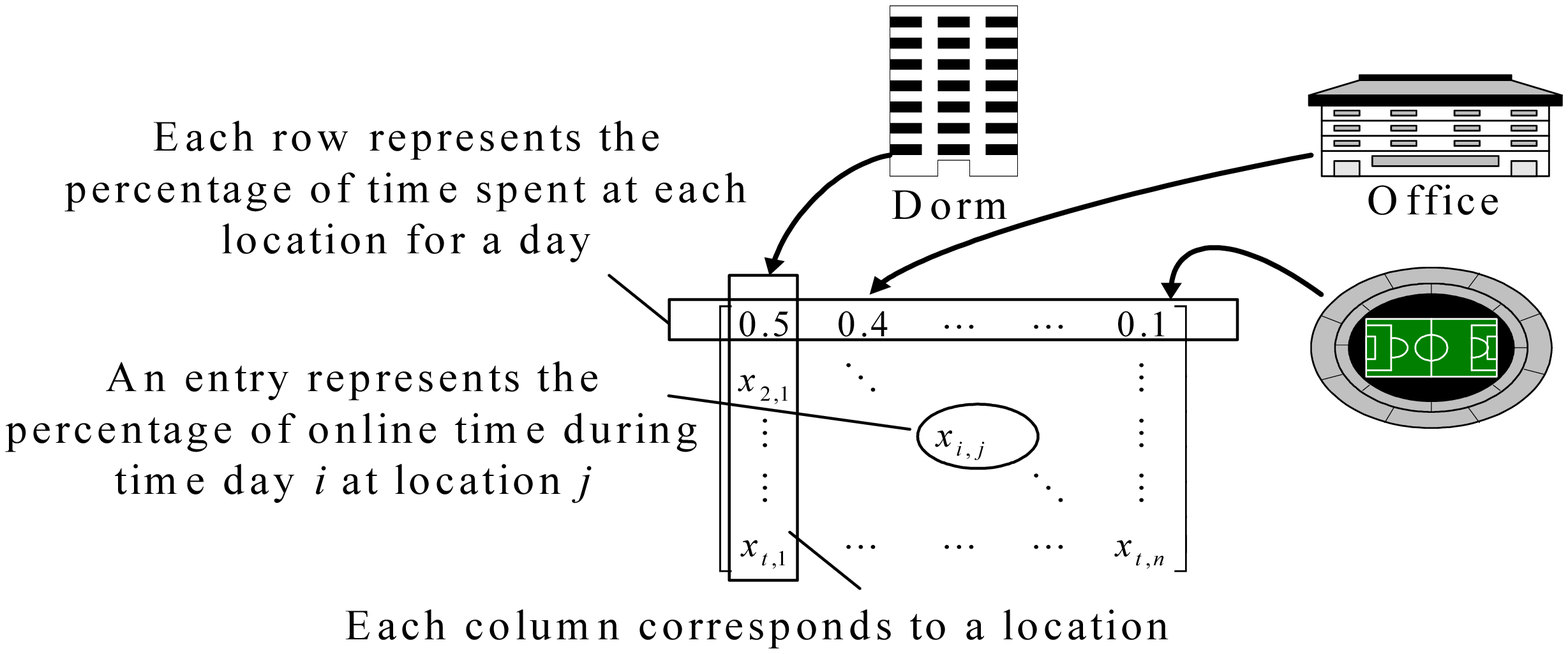} 
\caption{Illustration of the association matrix to describe a given user's location visiting preference.}
\label{matrix_illustration}
\end{figure}

We represent mobile user behavior of a given user using the {\it association matrix} as
illustrated in Fig. 1. In the matrix, each row vector describes the
percentage of time the user spends at each location on a day, reflecting the
importance of the locations to the user\footnote{While there may be numerous other representations of user
behavior, we shall show that this representation possesses desirable
characteristics for the purposes of this study. Further investigation of
other representations is a subject of future work.}. In \cite{MOBI07}
it has been shown that the {\it location visiting preferences} can be
leveraged to classify users of wireless networks on university
campuses. For a given user, the singular value
decomposition (SVD)~\cite{LP-norm} is applied to its {\it association matrix} $M$, such that
\begin{equation} \label{svd-eq}
M = U \cdot \Sigma \cdot V^T,
\end{equation}
where a set of {\it eigen-behavior} vectors, $v_1, v_2, ...,
v_{rank(V)}$ that summarize the important trends in the original 
matrix $M$ can be obtained from matrix $V$,
 with corresponding weights $w_{v_1}, w_{v_2}, ...,
w_{v_{rank(V)}}$ calculated from the eigen-values in matrix $\Sigma$. 
This set of vectors are referred to as the {\it
behavioral profile} of the particular user, denoted as $BP(M)$, as
they summarize the important trends in user $M$'s behavioral
pattern. The {\it behavioral similarity} metric between two users $A$
and $B$ is defined based on their behavioral profiles,
vectors $a_i$'s and $b_j$'s and the corresponding weights, as
\begin{equation} \label{sim-index}
\small
Sim(BP(A),BP(B)) = \sum^{rank(A)}_{i=1} \sum^{rank(B)}_{j=1}
w_{a_i}w_{b_j}\vert a_i \cdot b_j \vert,
\end{equation}
which is essentially the weighted cosine similarity between the two
sets of {\it eigen-behavior} vectors.

\subsection{Traces} \label{traces}

In this paper, we seek a realistic, deep understanding of user behavior patterns
by analyzing semester/quarter-long user behavioral logs collected from operational campus 
networks from public trace archives~\cite{MobiLib-web,CRAWDAD-web}.
We present results based on two data sets from
the University of Southern California (USC) and the Dartmouth College
(Dartmouth). The details of the data sets are listed in Table
\ref{trace-facts}.

\begin{table}
\caption{Facts about studied traces}
\label{trace-facts}
\begin{center}
\begin{tabular}{|c||c|c|}
\hline
Trace source & USC \cite{MobiLib-data} & Dartmouth \cite{Dart-movement-data} \\
\hline
Time/duration & 2006 spring & 2004 spring  \\
 of trace & semester & quarter \\
\hline
Start/End & 01/25/06- & 04/05/04- \\
time & 04/28/06 & 06/04/04 \\
\hline
Unique & \mr{137 buildings} & 545 APs/ \\
locations & & 162 buildings \\
\hline 
Unique MACs analyzed & 5,000 & 6,582 \\
\hline
\end{tabular}

\end{center}
\end{table}

We choose to use WLAN traces as they are the largest user behavioral
data sets available. The information available from these anonymized 
traces contains many aspects of the network usage (e.g., time-location information of the 
users by tracking the association and disassociation events with the access points,
amount of traffic sent/received, etc.). The richness in user behavioral
data poses a challenge in {\it representing} the user behavior in a 
meaningful way, such that the representation not only reveals an intrinsic, stable
behavioral profile of a user, but the identified behavioral profile
also leads to practical applications.
We show in this paper that the {\it location visiting preferences}
 (which is only a subset of the user behavioral data) is a 
stable attribute for both individual users and the relationship 
between users. This property will prove quite valuable to 
the design of efficient message dissemination
schemes, which we empirically validate using the above traces. 

\section{Understanding Spatio-Temporal Characteristics of User Behavioral Patterns} \label{sec:mobipattern}

In this section we introduce our analysis of user behavioral
patterns and its significance on the service design.
While previous works on user classification based on 
long-term behavioral trend~\cite{MOBI07, BU-classify, classify-Dart}
are useful and in line with our goal, the stability of such classification
over time has not been studied systematically.
In particular, the short-term behavior of a user may
deviate significantly from the {\it norm}, and the {\it stability} of user 
behavioral profiles is a decisive factor for whether it can be leveraged
to represent the user's future behavior. In this section we investigate the following
questions: (1) How long of behavioral
history do we need to classify a user? and (2) How much does the behavior
of a given user and its relationship with other users change
with respect to time?

\begin{figure}
\centering

\includegraphics[width=3.0in]{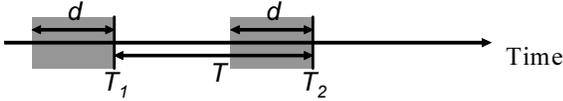} 

\caption{Illustration: consider the trailing $d$ days of behavioral profile
 at time points that are $T$ days apart.}
\label{time-scale}

\end{figure}

\begin{figure}
\centering
\includegraphics[width=2.5in]{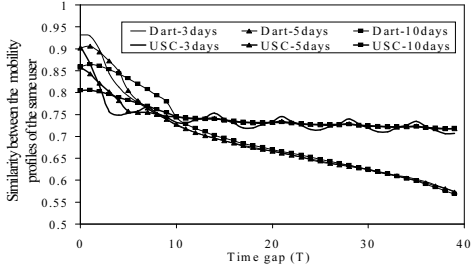} 
\caption{Similarity metrics for the same user at time gap $T$ apart.}
\label{self-stability}
\end{figure}

\begin{figure}
\centering

\includegraphics[width=2.5in]{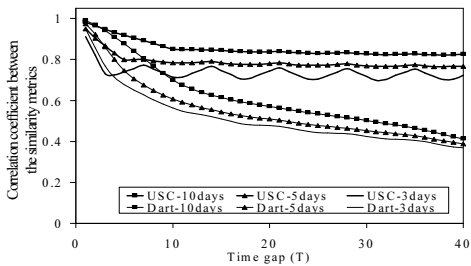} 

\caption{Correlation coefficient of the similarity metrics between the
same user pair at time gap $T$ apart.}

\label{relationship-stability}

\end{figure}

We consider the effect of the amount of past
history (of user behavior) on its {\it behavioral profiles}. Each
user uses the location visiting preference vectors in the past $d$
days to summarize the behavior in the most recent history -- the user
retains $d$ location visiting preference vectors for these days,
organize them in a matrix, and use singular value decomposition to
obtain the {\it behavioral profile}, as described in
section \ref{sec: classify}. We seek to understand how $d$ influences 
the representation and similarity calculations. More
specifically, we look into two important aspects: (1) Whether the
representation of a given user is stable across time, and (2) whether
the relationships between user pairs remain stable as time evolves.

We first consider the stability of the representation of a given
user. Considering two points in time that are $T$ days apart, we
obtain the {\it behavioral profiles} for the same user at both end
points, using the logs of the trailing $d$ days ending at those
end points, as illustrated in Fig. \ref{time-scale}. Then we use the similarity metric defined in
Eq. (\ref{sim-index}) to compare how stable a user's behavioral profile is to one's
former self after $T$ days has elapsed. The average results with
various values of the time gap, $T$, and considered behavioral history
$d$ are shown in Fig. \ref{self-stability}. We notice that, even
if we collect a short history of user behavior (say
$d=3$), the representation is similar to the behavior of the user for
a long time into the future. When we consider $T=35$ days apart, the
behavioral profiles from the same user still show high
similarity, at about $0.6$. The amount of history used does not influence
the result too much when the considered $T$ is large enough to 
avoid overlaps in the used behavioral history (i.e., when $T > d$). We conclude
that on university campuses, the {\it behavioral profile} for a given user is
stable, i.e., it remains highly similar for the same user across time.
One interesting note is that, when the behavioral profile includes
only part of a week ($d < 7$), the similarity of the user to its former self
shows a weekly pattern (i.e., when $T$ is an integer multiple of seven, the similarity
peaks), especially in USC.


Second, we try to quantify how the behavioral similarity  between the same pair
of users varies with time. For this part, we use Eq. (\ref{sim-index})
to calculate the similarity between two users, $A$ and $B$, at two 
points in time, $Sim_{T_1}(A,B)$ and $Sim_{T_2}(A,B)$, where $T_1$ 
and $T_2$ are $T$ days apart. We perform this calculation to
all user pairs, and then calculate the correlation coefficient of the 
similarity metrics obtained after a $T$-day interval, as
\begin{equation} \label{correlation_coef}
\small
r = \frac{\sum_{\forall {A,B}}( X - \overline{X})(Y - \overline{Y}) }{N S_{X}S_{Y}},
\end{equation}
where $X = Sim_{T_1}(A,B)$ and $Y = Sim_{T_2}(A,B)$, and the notations
$\overline{X}$ and $S_X$ denote the average and standard deviation of $X$, 
respectively. $N$ is the total number of user pairs. The correlation coefficient quantifies
how stable the relationship between user pairs is. We repeat the
calculation for all pairs of users with various $d$ and $T$ values to
arrive at Fig. \ref{relationship-stability}. We observe that the
similarity metrics between user pairs correlate reasonably well if the
considered time periods are not far apart. For $T$ smaller than one
week, the correlation coefficient is above $0.62$. This indicates,
once the similarity between a pair of user is obtained, it remains a
reasonable predictor for their mutual relationship for some time
period into the future. Although the reliability of the stale
similarity data decreases with respect to time, the current similarity of a user
pair remains moderately correlated to their future similarity, in the
time range up to several weeks. The correlation is above $0.4$ for
up to five weeks.



\textbf{The investigation establishes that the user behavioral profile is a
stable feature to represent the users -- the representation of an
individual user and the relationship between users are well
correlated with the past history for the near future.} Thus we map the behavioral profile to
a virtual {\it behavioral space}~\cite{M-space-routing}, in which each
user's behavior is quantified as a high dimensional point\footnote{The dimension of 
the behavioral space is the same as the {\it mobility preference vector} representation,
typically in the order of a hundred for these two campuses.}. The mutual
similarity metric between users is a function of their respective
positions in this space. In this paper, when we say two users are {\it
similar}, it means they are {\it close} in the behavioral space (i.e.,
the {\it distance} between the two users is small). We also use
the term {\it neighborhood of a node} to refer to
the other nodes that are {\it similar} to this particular node in the
behavioral space.

\section{The Behavior-driven Communication Paradigm} \label{sec:transition}

Profiling users based on stable behaviors is a fundamental
step to understand human behavior.
Motivated by the stability of user behavioral profiles, we introduce
a {\it behavior-driven communication paradigm} where we use
{\it user behavioral profiles}, instead of
network IDs, to represent users.
We envision that such a radical approach has several benefits.

First, it enables behavior-aware message delivery in the network
without mapping attributes to
network IDs.  As each user maintains its behavioral profile, it is now
possible to deliver announcements about a sports event on campus
towards sports enthusiasts (e.g., people who visit the gym often) or
advertise a performance at the school auditorium to the regular
attendees of such events.

Second, it facilitates the discovery of nodes with certain behavior
patterns. Consider, for example, in the message
ferry~\cite{ferry} architecture where nodes with high mobility move
messages across the network to facilitate the communication between
otherwise disconnected nodes. One can choose a target profile that reflects a
mobility profile and thus eliminate the need of knowing
the identity of the ferry beforehand or enforcing this mobility
pattern on a controlled node -- a typical user who happens to have the
desired mobility pattern can be discovered and serve as a ferry.

Our
{\it behavior-driven communication paradigm} is applicable to several
architectures. In the {\it centralized
server-based architecture}, user profiles could be collected and stored at a
data repository, and mined for user classification, abnormality detection,
or targeted advertisements. In the {\it cellular networks},
the low-bandwidth channel between the users and the infrastructure can
be leveraged to exchange behavioral profiles and match users.
In this paper, however, we consider a {\it decentralized
infrastructure-less networks}, and focus on how stable behavioral profiles
are used for better message dissemination.
We name this scheme as {\it CSI},
since it is a {\it C}ommunication scheme based on
the {\it S}table, {\it I}mplicit structure in human networks.


\section{Protocol Design} \label{sec:proto}

In this section, we first present our premises and design requirements
for the CSI schemes. We then discuss the design of the CSI schemes
based on in-depth understanding of the relationship between similar
behavioral profiles and encounter events.

\subsection{Assumptions and Design Requirements} \label{sec:design_goal}

We assume that each node profiles {\it its own behavioral pattern} by
keeping track of the visiting durations of different
locations and summarizing the behavioral
profile using the technique discussed in \ref{sec: classify}. This is
an individual effort by each node involving no inter-node
interactions. This can be done by the nodes over-hearing the beacon
signals from the fixed access points in the environment to find out
its current location. Note that, the use of these beacon signals is
only for the node to profile its own behavior -- they are not used to
help the communication in our protocols (we will re-visit detailed
points of this assumption in section \ref{sec:disc}). Also, for the ease of
understanding, we assume in this section that nodes are willing to
send its behavioral profiles to other nodes when needed. A 
privacy-preserving option that eliminates this operation
is also discussed in section \ref{sec:disc}.

The goal of our {\it CSI} scheme is to reach a group of nodes matching
with the target profile specified by the sender, under the following 
performance requirements: (1) The protocol should be scalable, in particular
not being dependent on a centralized directory to map target profiles to 
user identities. (2) It should work in an efficient manner and avoid
transmission and storage overhead when possible. Also, it should
avoid control message exchanges in the absence of data traffic.
 (3) The syntax of the
target profile should be flexible, allowing the target profile
to be not in the same context as the behavioral profiles we use to
represent the users. Also the operation of the 
protocol should be flexible to allow tradeoff between various performance
metrics. And finally, (4) the design should be robust and help in
protecting user privacy.

We design two modes of operation for
the {\it CSI} scheme under the above requirements. When the target profile is in the same 
context as the behavioral profile (in our example, since the behavioral 
profile is a summary of user mobility, this corresponds to the scenario when the 
target profile describes users that {\it move} in a particular way), the 
{\it CSI:Target mode (CSI:T)} should be used. When the target profile is irrelevant
to the behavioral profile (e.g., when I want to send to everyone interested
in movies on campus), the {\it CSI:D mode} should be used
instead. Although it seems that the applicability of {\it CSI:T} is limited,
we note that the behavioral profile (in terms mobility) can sometimes 
be used to infer other social aspects of the users, such as affiliations
or even interests (e.g., people who visit the gym often should like sports in general).
Such inferences expand the scenarios in which {\it CSI:T} can be used.
When this is not possible, {\it CSI:Dissemination mode (CSI:D)} provides a more generic option.

The major challenge involved in the design process is that each node
is only aware of the behavioral profile of itself. Furthermore,
we require no persistent control message
exchanges for the nodes to ``learn'' the structure of the network
proactively when they have no message to send.
Nodes only compare their behavioral profiles {\it when they are involved in
message dissemination}.
Based on this very limited knowledge about the behavioral space, a node must predict
how useful a given encounter opportunity is in
terms of achieving the fore-mentioned requirements. Since encounter events
may occur sporadically in sparse, opportunistic networks, the nodes
must make this decision for each encounter event independent of 
other encounter events (that may occur long before or after the current
one under consideration). Such a heuristic must
rely on the understanding of the relationship between nodal behavioral
profiles and encounters, which we discuss the next.

\subsection{Relationship between Behavioral Profiles and Encounters} \label{sec:mobi_vs_enc}

\begin{figure*}
\begin{minipage}[t]{2.3in}
\centering

\includegraphics[width=2.0in]{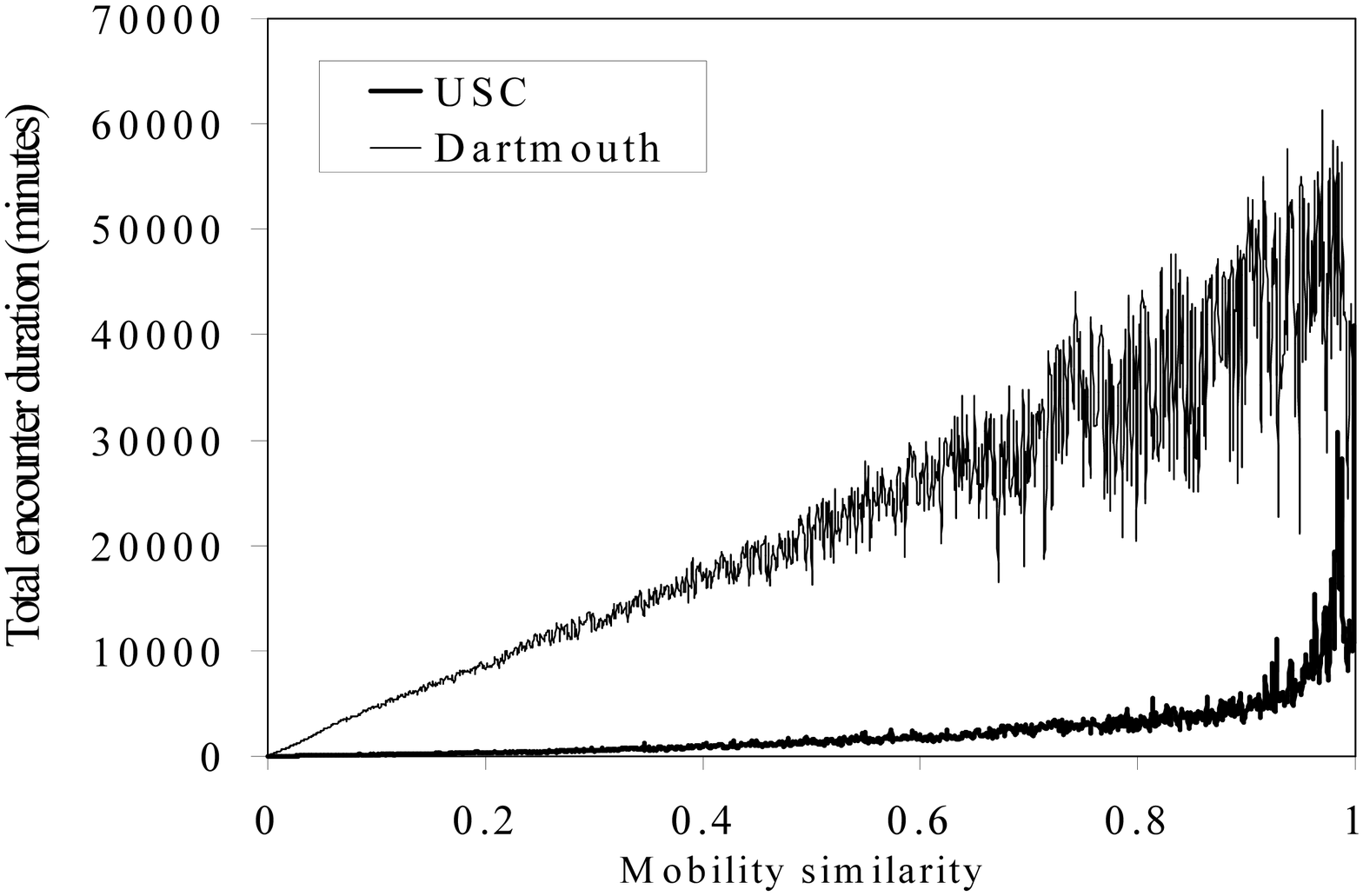}

\footnotesize{(a) Total encounter duration.}
\hfill
\end{minipage}
\begin{minipage}[t]{2.3in}
\centering

\includegraphics[width=2.0in]{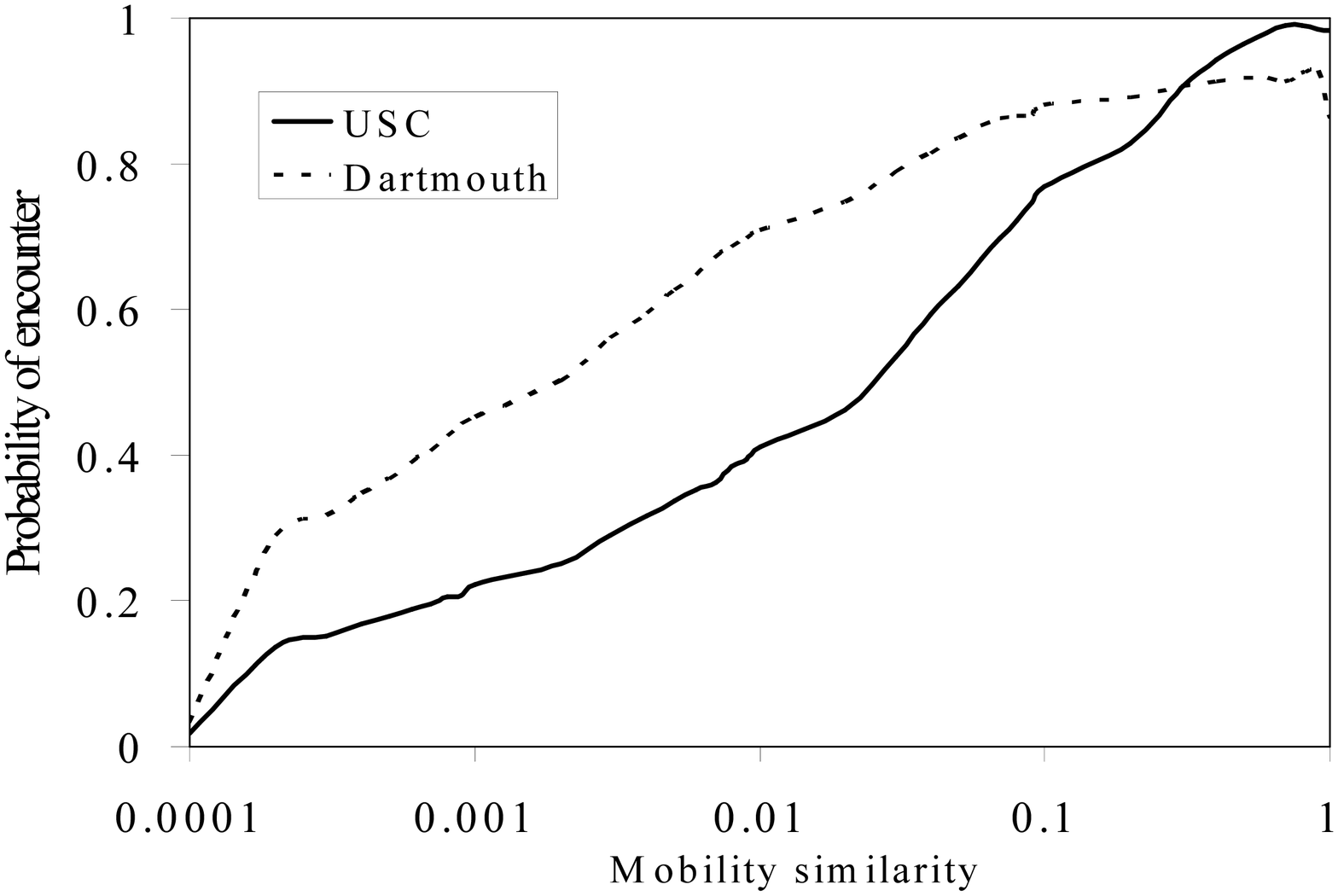}

\footnotesize{(b) Encounter probability.}

\end{minipage}
\hfill
\begin{minipage}[t]{2.3in}
\centering

\includegraphics[width=2.0in]{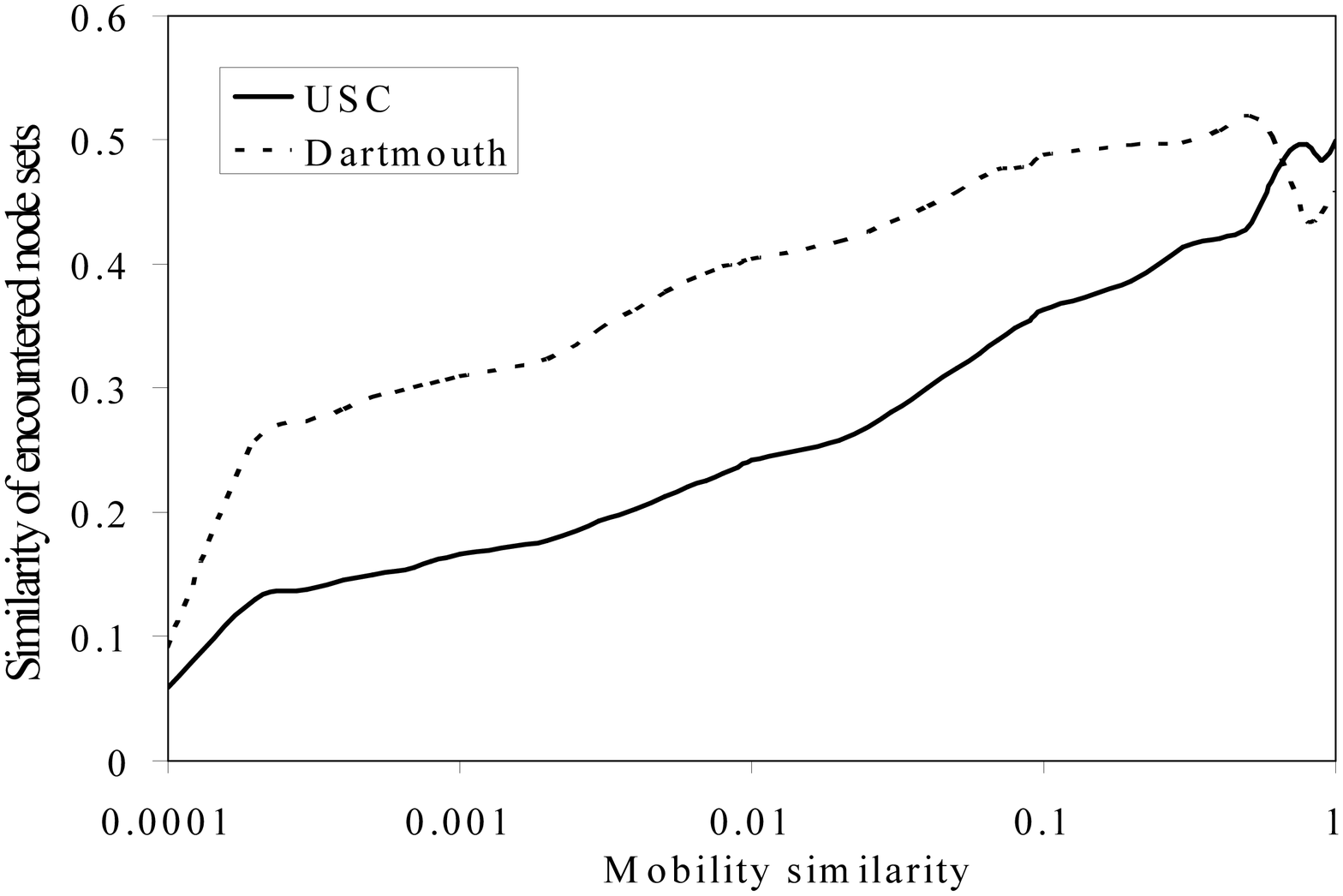}

\footnotesize{(c) Similarity of encountered node sets.}

\end{minipage}
\hfill

\caption{Relationship between the similarity in behavioral pattern and
other quantities.}

\label{Msim_vs_X}
\end{figure*}

We now analyze the relationship between user behavioral
profiles and a key event for user-to-user communication
in an infrastructure-less network -- {\it encounters}. {\it
Encounters} in mobile networks refer to events when users
are within the radio range of each other and direct communication
between the involved devices is possible. In this paper, based on the
WLAN traces, we assume that when two users visit the {\it same
location} during overlapped time intervals, they {\it encounter} with
each other.

While it seems intuitive that users visiting similar locations should
encounter with each other with higher probability, this is {\it not
obvious} on university campuses. Students and faculty have their
own schedules, and they may rarely encounter due to the difference in
their schedules although they might be in the same building at different times.
Hence we investigate the relationship between
behavioral profiles and encounter events, first as a sanity check of
our intuition, and more importantly, to understand the
relationship between the behavioral patterns and various aspects of
the encounter events (e.g., the encounter probabilities, encounter
durations, etc.). This helps reveal the {\it
implicit structure} existing in mobile human networks, which is the
key to the design of the {\it CSI} schemes in the following sections.

We classify all node pairs into different bins of behavioral
similarity metric (as defined in Eq. (\ref{sim-index})), and obtain
various characteristics of encounter events as a function of the
pair-wise behavioral similarity. In Fig. \ref{Msim_vs_X} (a), we show
the aggregate encounter time duration between an average pair of nodes
given the behavioral similarity. In Fig. \ref{Msim_vs_X} (b), we show
the probability for a given node pair to encounter with each other,
given their similarity. Combining these two graphs, we see that
\textbf{if two users are similar in behavioral profiles, they are much
more likely to encounter, and the total time they encounter with each
other is much longer -- an indication that nodes with
{\it similar behavioral profiles} indeed are more likely to
have better opportunities to communicate.} When two users are similar enough (with behavioral
similarity larger than $0.3$), they are almost guaranteed to encounter
at some point (with probability above $0.9$). However, we note that
some ``random'' encounter events happen
between dissimilar users. For users with very low (almost zero)
similarity, the probability for them to encounter is not zero,
although such encounter events are much less reliable (i.e., they
occur with much shorter durations, see Fig. \ref{Msim_vs_X} (a)).

In Fig. \ref{Msim_vs_X} (c) we further compare the behavioral
similarity of node $A$ and $B$ versus the sets of nodes $A$ and $B$
encounter. We denote the set of nodes $A$ encounters with as
$E(A)$. The similarity of the two sets of nodes is quantified by $ |
E(A) \cap E(B) | / | E(A) \cup E(B) |$, where $| \cdot |$ is the
cardinality of the set. This graph shows, \textbf{as two nodes are
increasingly similar, there is larger intersection of nodes they encounter.
When an unlikely
encounter event between {\it dissimilar nodes} occurs, it helps both
nodes to gain access to a very different set of nodes, which
they are unlikely to encounter directly.}

The above findings relate to the SmallWorld encounter patterns
between mobile users~\cite{group-study}. The key features of SmallWorld
networks~\cite{smallworld} are high clustering coefficient and low average
path length. In the human networks we analyze in this section,
people with similar behavior
form ``cliques''. The ``random'' encounter events between
dissimilar nodes build {\em short-cuts} between these cliques to shorten
the distances between any two nodes. We leverage these properties
in the protocol design.

\subsection{CSI:Target Mode} \label{sec:MPC}

In the {\it CSI:target mode (CSI:T)}, the sender specifies the {\it
target profile (TP)} for the recipients which must have the same format
and semantics as that of  the user behavioral profile, i.e., in our case
 the {\it TP} is a summarized {\it mobility preference} vector (i.e.,
the percentage of times the target node(s) visit various
locations). For example, we could reach people who like
sports by  sending messages to those who visit the gym
regularly. This criteria could be set up by specifying the {\it TP} as
a vector with only one $1$ corresponding to the gym location (hence
only time spent at this location is considered).  If a given user $A$
has $Sim(BP(A), TP) > th_{sim}$, i.e., its behavioral profile,
$BP(A)$, is more similar to $TP$ than a sender specified threshold, we
say node $A$ belongs to the group of {\it intended receivers}.  This
threshold is set by the sender according to the desired degree of
similarity to the $TP$. The $TP$ and the
threshold, $th_{sim}$, are included in the message header to describe
the intended receivers of the message.

\begin{figure}

\centering

\includegraphics[width=1.8in]{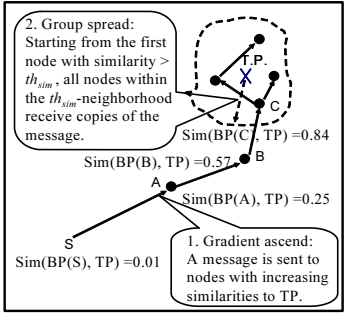}

\caption{Illustration of the CSI:T scheme in the {\it high dimension behavioral space}. One copy of the message follows increasing similarity gradient to reach the neighborhood of the target profile, then triggers group spread.}
\label{MPC_illustration}
\end{figure}

We first discuss the intuition behind the design of the {\it CSI:T
mode} using Fig. \ref{MPC_illustration} as an illustration. As per
section \ref{sec:mobi_vs_enc}, to deliver messages to receivers
defined by a given {\it TP}, one way is to gradually move the message
towards nodes with increasing {\it similarity} to the {\it TP} via
encounters, in the hope that such transmissions will improve the
probability of encountering the intended receivers. Finally, when the
message reaches a node {\it close} to the {\it TP} (in the behavioral
space), most nodes encounter frequently with this node are also
similar to {\it TP}. Hence, the message should be spread to other
nodes in the {\it neighborhood} (in the behavioral space) of the node.

Consider the pseudo-code in Algorithm \ref{MPC-algo}. There are two
phases in the operation, the {\it gradient ascend phase} and the {\it
group spread phase}. (1) Starting from the sender, if node $A$
currently holding the message is not an intended receiver (i.e., $Sim(
BP(A), TP ) < th_{sim}$), it works in the {\it gradient ascend phase},
otherwise it works in the {\it group spread phase}.  (2) In the {\it
gradient ascend phase}, for each encountered node, the current message
holder asks the behavioral profile of the other node, and if the other
node is more similar to the $TP$ in the behavioral space, the
responsibility of forwarding the message is passed to this node.  One
can imagine that these similarities form an inherent {\it gradient}
for the message to follow and reach the close neighborhood of the {\it
TP} in the behavioral space, hence the name {\it gradient ascend
phase}.  Note that, up to this point, there is only one copy of the
message in the network -- these intermediate nodes who are not similar
to the $TP$ only forward the message once.  (3) When the message
reaches a node with similarity larger than $th_{sim}$ to the {\it TP},
the {\it group spread phase} starts. This intended receiver holds on
to the message, and requests the behavioral profiles from nodes it
encounters. If they are also intended receivers, copies of the
messages will be delivered to them. All intended receivers, after
getting the message, continue to work in the {\it group spread
phase}. Although multiple copies of the message are generated in the
{\it group spread phase}, it is triggered only when the message is
close to the {\it TP}, thus most of the encounter events and inquiries
will occur among the {\it intended receivers}, reducing unnecessary
overhead.

\begin{algorithm}
\SetVline
\SetCommentSty{textbf}
\tcc{$BP(A)$: {\it Behavioral profile} of node $A$}
\If{node $A$ has the message}{
\eIf{ $Sim( BP(A), TP ) > th_{sim}$}{
       Initiate {\it Group\_spread}()\;
}{
       Initiate {\it Gradient\_ascend}()\;
}
}

{\it Gradient\_ascend}()\{

  \While{the message is not sent}{
       \ForEach{node $E$ encountered}{
           Get $BP(E)$ from $E$\;
           \If{$Sim( BP(E), TP ) > Sim( BP(A), TP )$}{
               Send message to $E$\;}
        }
}
\}

{\it Group\_spread}()\{

       \ForEach{node $E$ encountered}{
           Get $BP(E)$ from $E$\;
           \If{$Sim( BP(E), TP ) > th_{sim} $}{Send message to $E$\;}
        }
\}
\caption{Algorithm for the CSI:T mode} \label{MPC-algo}
\end{algorithm}

\subsection{CSI: Dissemination Mode} \label{sec:MIPC}

\begin{figure}

\centering
\includegraphics[width=2.0in]{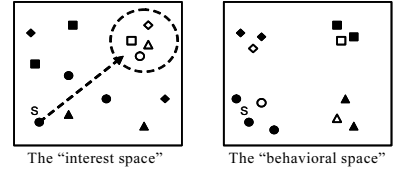}
\caption{Illustrations of the {\it CSI:D} scheme. Left chart: The goal is to send a message to a group of nodes with a similar characteristic in the {\it interest space} (white nodes in the circle). Right chart: However, they may not be similar to each other in the behavioral space (nodes with the same legend represent similar nodes in the behavioral space).}
\label{I-B-space}

\end{figure}

In the {\it CSI:Dissemination mode (CSI:D)}, there does not exist a direct
relationship between the target profiles of the recipients and their
measured behavioral profiles. One particular example is to reach
people who like movies on campus. If there is no movie theaters
on campus, the measured behavioral profiles (i.e., mobility preference)
cannot be used to infer such an interest. This situation is illustrated in
Fig. \ref{I-B-space}. It appears there is little insight
provided by the similarities between the nodal behavioral profiles to
guide message propagation, as the intended receivers in this case
may be scattered in the behavioral space, and the relationship
between the target profile and the behavioral profile cannot be
quantified. Although it is always possible to
reach most users through epidemic routing, this leads
to high overhead, and requires all nodes in the network to keep a copy
of the message. The objective of {\it CSI:D mode} is to
reduce the numbers of message copies transmitted and stored in the
network, yet make it possible for most nodes to get a copy quickly,
if they belong to the intended receivers.

\begin{figure}

\centering

\includegraphics[width=2.2in]{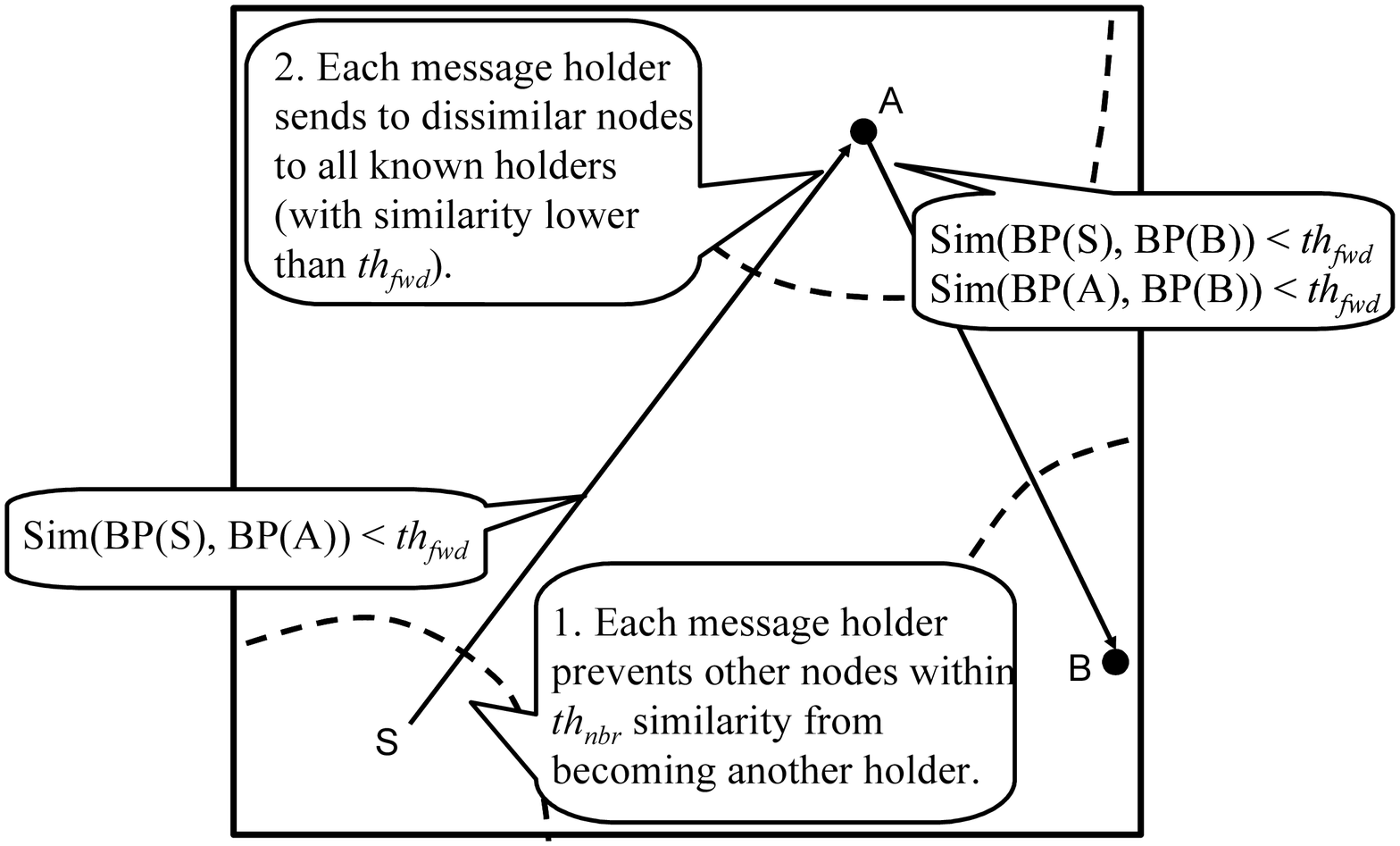} 

\caption{Illustration of the CSI:D scheme. The idea is to select the message holders in a non-overlapping fashion to cover the entire behavioral space.}
\label{MIPC_illustration}
\end{figure}

We again first discuss the  intuition behind the design
of the {\it CSI:D mode} in this paragraph, using Fig. \ref{MIPC_illustration} as an
illustration. From section \ref{sec:mobi_vs_enc}, \textbf{since the
nodes with high similarity in their behavioral profiles are almost
guaranteed to encounter, there is really no need for each of them to
keep a copy and disseminate the message. Electing a few {\it message
holders} within a single group of similar nodes would suffice.}
This intuition leads to the construction of our message dissemination
strategy for the {\it CSI:D}. We aim to have only one
{\it message holder} among the nodes who are similar in their
behavioral profiles (or equivalently, pick only one {\it
message holder} within a {\it neighborhood} in the behavioral
space. In Fig. \ref{I-B-space}, this corresponds to having only one
message holder from each group of nodes with the same legend).
We add the messages holders carefully to avoid overlaps in the
encountered nodes among message holders. As suggested by
Fig. \ref{Msim_vs_X} (c), we should \textbf{select nodes that are very {\it
dissimilar} in their behavioral profiles to achieve low overlaps.} Recall
that dissimilar node pairs still encounter with non-zero probability,
our design philosophy is to leverage these ``random'' encounter events
as {\it short-cuts} to navigate through the behavioral space
efficiently, hopping across the space to reach dissimilar nodes with
relatively few message transmissions. Such a design philosophy is
also related to the SmallWorld human network structure -- a message
will be received by an intended receiver shortly once it has reached
someone in the receiver's ``clique''.

Consider the pseudo-code in Algorithm \ref{MPC-algo}.
(1) The sender itself starts as the first message
holder in the network. (2) Each message holder tries to strategically
add additional message holders in the network. When it encounters
with other nodes, it asks for the behavioral profile of the other node
to be considered as a potential additional message holder.
Each message holder keeps a list of the behavioral profiles of all known
message holders\footnote{Note this list does not necessarily contain
all holders in the network. Message holders that are added by a
particular message holder are not known to other holders until
they meet and sync the lists.}, and the new node has to be dissimilar
(with the similarity metric lower than a threshold, $th_{fwd}$) to all
known holders to be added as
a new message holder and keep another full copy of the message.
(3) If, on the other hand, this node is similar to the message holder
(i.e., within similarity threshold $th_{nbr}$), it uses a single bit to remember
that there is a message holder in its neighborhood and propagates this information to similar
nodes. This bit is used to prevent excessive message holders in the
same neighborhood, even if some nodes have not encountered with the
message holders directly.
(4) When holders encounter, they update
each other with the behavioral profiles of the known holders list, to
gain a better view of the situation of message spreading. (5) If
two similar holders encounter, one of them should cease to be a holder
to reduce duplicated efforts.

Each message holder is responsible for disseminating the actual
message to the intended receivers. The message holders
sends the {\it TP} specified by the sender in the message
to the encountered nodes. If the encountered node is an intended receiver, the
full message will be transferred.

\begin{algorithm}
\SetVline
\SetCommentSty{textbf}
\tcc{$BP(A)$: {\it Behavioral profile} of node $A$}
\tcc{$H_i(A)$: The $i$-th known holder of node $A$}
\tcc{$holder\_in\_group(A)$: If $A$ knows there is a message holder in its neighborhood}
\uIf{node $A$ is a message holder}{
   \ForEach{node $E$ encountered}{
     Get $BP(E)$\;
      \eIf{$E$ is not a holder}{ 
               \uIf{$Sim( BP(E), BP(H_i(A)) ) < th_{fwd} \forall i$ and $holder\_in\_group(E) = false$}{ 
                  Elect $E$ as an holder\;
                  Add $BP(E)$ to holder list\;
                  Send the message\; 
                  Send $BP(H_i(A)), \forall i$\;
               }
               \ElseIf{$Sim( BP(E), BP(H_i(A)) ) > th_{nbr}$ for any $i$}{
                 Let $E$ set $holder\_in\_group(E) = true$\;
               }
      }{
        \eIf{$Sim( BP(E), BP(A)) > th_{nbr}$}{
              $A$ ceases to be a holder\;
        }{
           Sync holder lists between node $A$ and $E$\;
        }
      } 
  }
}
\ElseIf{$holder\_in\_group(A) = true$}{
      \ForEach{node $E$ encountered}{
         Get $BP(E)$\;
         \If{$Sim (BP(A), BP(E)) > th_{nbr}$}{
             Let $E$ set $holder\_in\_group(E) = true$\;
         }
      }
}
\caption{Algorithm for CSI:D mode.} \label{MIPC-algo}
\end{algorithm}

\section{Simulation Results} \label{sec:exp}

In this section, we perform extensive simulations with the CSI
schemes, based on the derived encounters between users
from the two empirical traces. We compare the performances of our
proposal to oracle-based forwarding decisions to show that our
performance is close to the optimum (in terms of the delivery success 
rate and the overhead), and does not fall much behind in delay.
We also compare CSI to epidemic routing~\cite{epidemic}
and variants of random walk\footnote{The CSI could not be directly compared with existing
routing schemes (e.g., \cite{PRoPHET, M-space-routing, socialnet-routing, group-routing})
in DTN as most of them have a different routing objective: reaching a particular
network ID.}. In all the simulation cases, we split the traces into two
halves, use the first half to obtain the behavioral profiles for all
users, and then use the second half of the trace to evaluate the
success of our proposed schemes.

\subsection{CSI:Target Mode}

\subsubsection{Simulation Setup}

In the scenario of CSI:T mode, the sender specifies the {\it TP} and a threshold of similarity $th_{sim}$. If a node shows a similarity metric higher than $th_{sim}$ to the {\it TP}, it is an intended receiver. In our evaluation, we use the top-$10$ dominant behavioral profile\footnote{We have also experimented with other target profiles, such as rarely visited locations on campuses or profiles that contain a combination of several locations, and the results are similar to those presented in this section.} (i.e., the behavioral profiles with the most number of people following it, typically in the order of hundreds) in our traces as the {\it TP}, and for each {\it TP} we randomly pick $100$ users as the senders generating messages targeting at the {\it TP}. We use the threshold $th_{sim} = 0.8$ as the transition point between the {\it gradient ascend phase} and the {\it group spread phase}.

We compare our {\it CSI:T} scheme with several other protocols discussed below. The {\it  epidemic routing}~\cite{epidemic} is a message dissemination scheme with simplistic decision rules: all nodes in the network send copies of messages to all the encountered nodes who have not received the message yet. The {\it random walk (RW)} protocol generates several copies of the message from the sender, and each copy is transferred among the nodes in a random fashion, until the hop count reaches a pre-set $TTL$ value. {\it Group spread only} is a simplified version of our protocol. It uses only the {\it group spread phase}, i.e., the original sender holds on to the message until it encounters with someone who is more similar than $th_{sim}$ to the {\it TP} and starts the {\it group spread phase} directly from there.

We also consider two protocols that require global knowledge of the future. The {\it optimal} protocol sends copies of the message only to the nodes which lead to the fastest delivery to the targeted receivers, and no one else. This is the oracle-based optimal protocol achievable if one has perfect knowledge of the future, and serves as the upper bound for performance. The {\it optimal single-forwarding-path} is the oracle-based protocol to find the fastest path to deliver the message to the neighborhood of the {\it TP} -- Using the knowledge of the future, it identifies the path that leads to the earliest message delivery to one of the intended receivers. Once a copy of the message is delivered to the  $th_{sim}$-neighborhood to the {\it TP}, it follows the same {\it group spread phase} as in CSI:T. This is the optimal performance (upper bound) for the family of protocols delivering one copy of message to the neighborhood of the target profile, if one chooses a good (shortest delay) path -- note that this shortest-delay path may not always follow an increasing gradient of similarities to the {\it TP}.

We compare these message dissemination schemes with respect to three important performance metrics: {\it delivery ratio}, {\it average delay}, and {\it transmission overhead}. The {\it delivery ratio} is defined as the percentage of the intended receivers (those with similarity greater than $th_{sim}$ to the $TP$) actually received the message. We account for the transmission overhead as {\it the total number of messages sent} in the process of delivery. See more discussions on the additional overhead of exchanging the behavioral profiles later in section \ref{sec:overhead}.

\subsubsection{Simulation Results}

\begin{figure}
\centering

\includegraphics[width=2.8in]{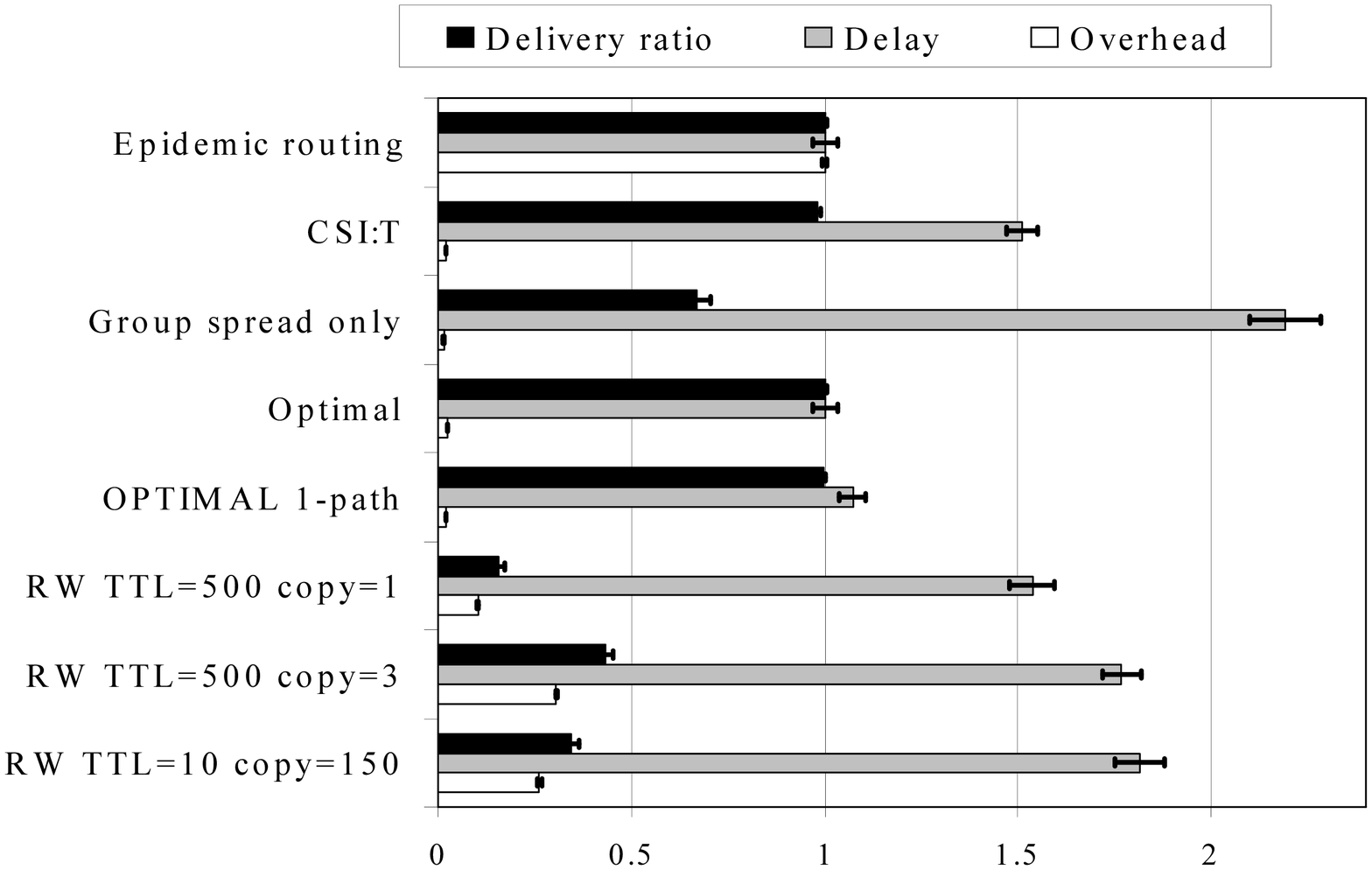} 

\footnotesize{(a) USC.}

\includegraphics[width=2.8in]{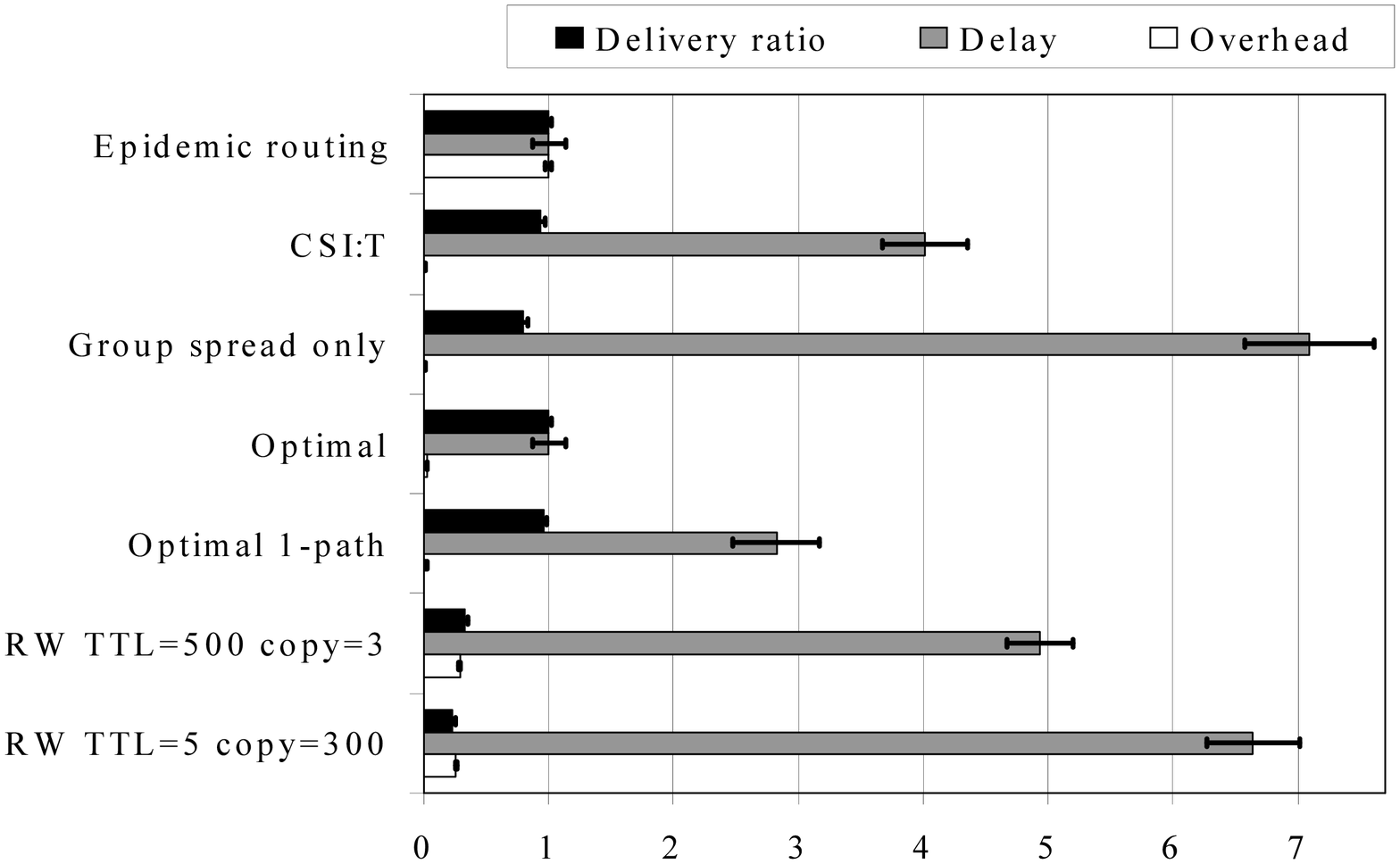} 

\footnotesize{(b) Dartmouth.}

\caption{Performance comparison of CSI:T to other protocols.}

\label{MPC_performance}
\end{figure}

We show the normalized performance metrics with respect to that of {\it  epidemic routing} (the relative performance for each protocol assuming {\it epidemic routing} is $1.0$) and its $95\%$ confidence intervals in Fig. \ref{MPC_performance}. We observe that {\it epidemic routing} leads to the highest overhead while its aggressiveness also results in the highest possible delivery ratio and the lowest possible delay. The {\it random walks} do not work well regardless the number of copies and the value of $TTL$, as they use no information to guide the propagation of the message towards the right direction. Our {\it CSI:T} protocol leads to a success rate close to the {\it epidemic routing} ($0.96$ for USC, $0.94$ for Dartmouth) with very small overhead ($0.02$ for USC, $0.018$ for Dartmouth). For the simplified version, {\it group spread only}, the delay is longer and the success rate is lower than our protocol. We will further investigate this phenomenon later.

When comparing {\it CSI:T} with the protocols with future knowledge, we see that there is really not much room for improvement in terms of the success rate and the overhead. Our gradient ascend approach in {\it CSI:T} is similar to what is achievable even one has the knowledge of the future in these two aspects. Specifically, {\it CSI:T} has more than $94\%$ of delivery rate and uses {\it less than $84\%$} overhead of the {\it optimal} strategy. The delay, on the other hand, has some room for improvement. Our gradient ascend phase generates only one copy of message from the sender and it moves towards the $TP$ following strictly ascending similarity. Comparing with the best (fastest) path to the $TP$ used in the {\it optimal single-forwarding-path}, our {\it CSI:T} has $1.40$ and $1.47$ times more delay, for USC and Dartmouth, respectively. If we compare with the {\it optimal} strategy, where multiple copies are generated whenever it helps to improve the delay, the difference is even larger. This calls for a further investigation of selecting good path(s) from the sender to the $TP$, which we leave out for future work.

\begin{figure}
\centering

\includegraphics[width=2.5in]{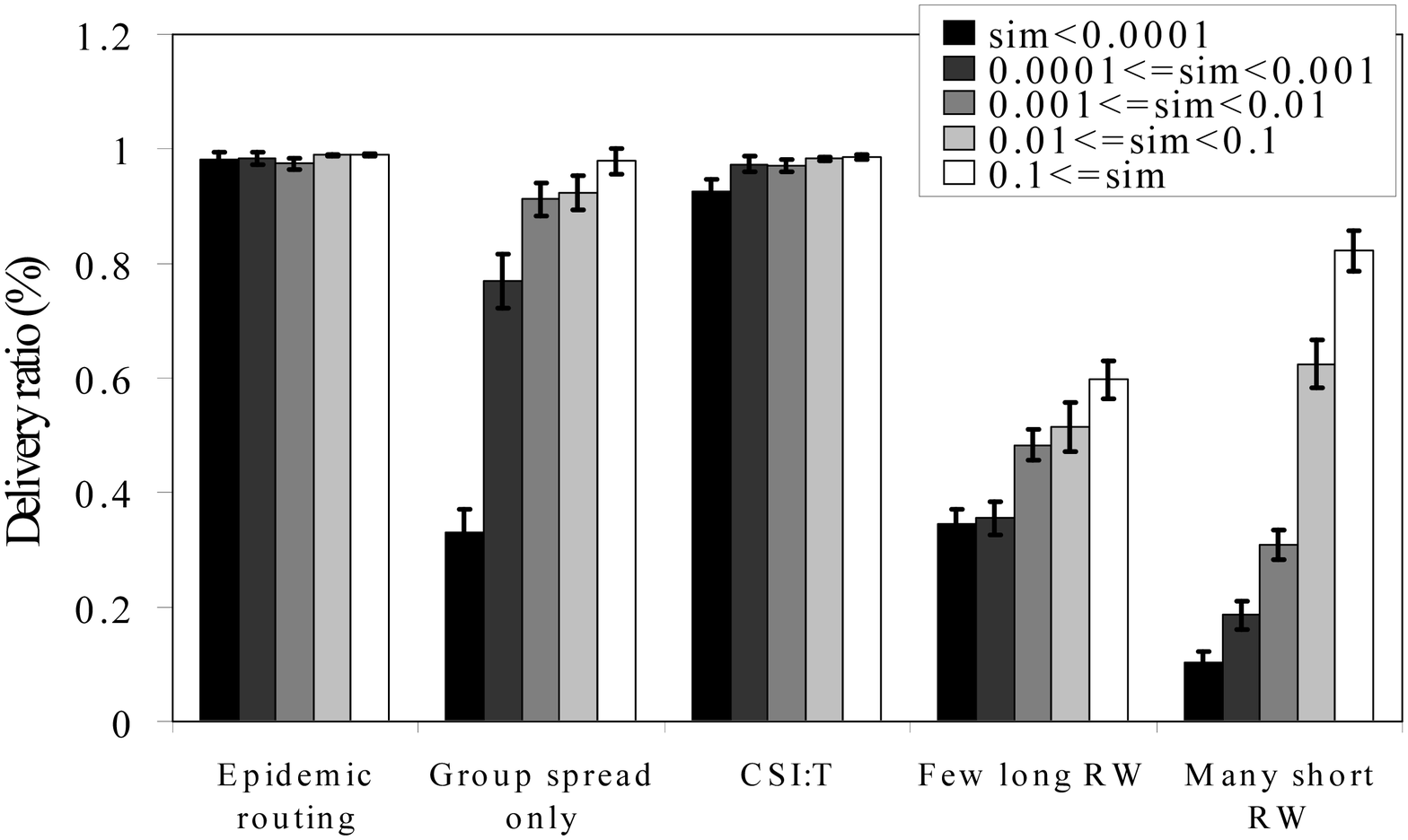} 

\footnotesize{(a) Delivery ratio.}

\includegraphics[width=2.5in]{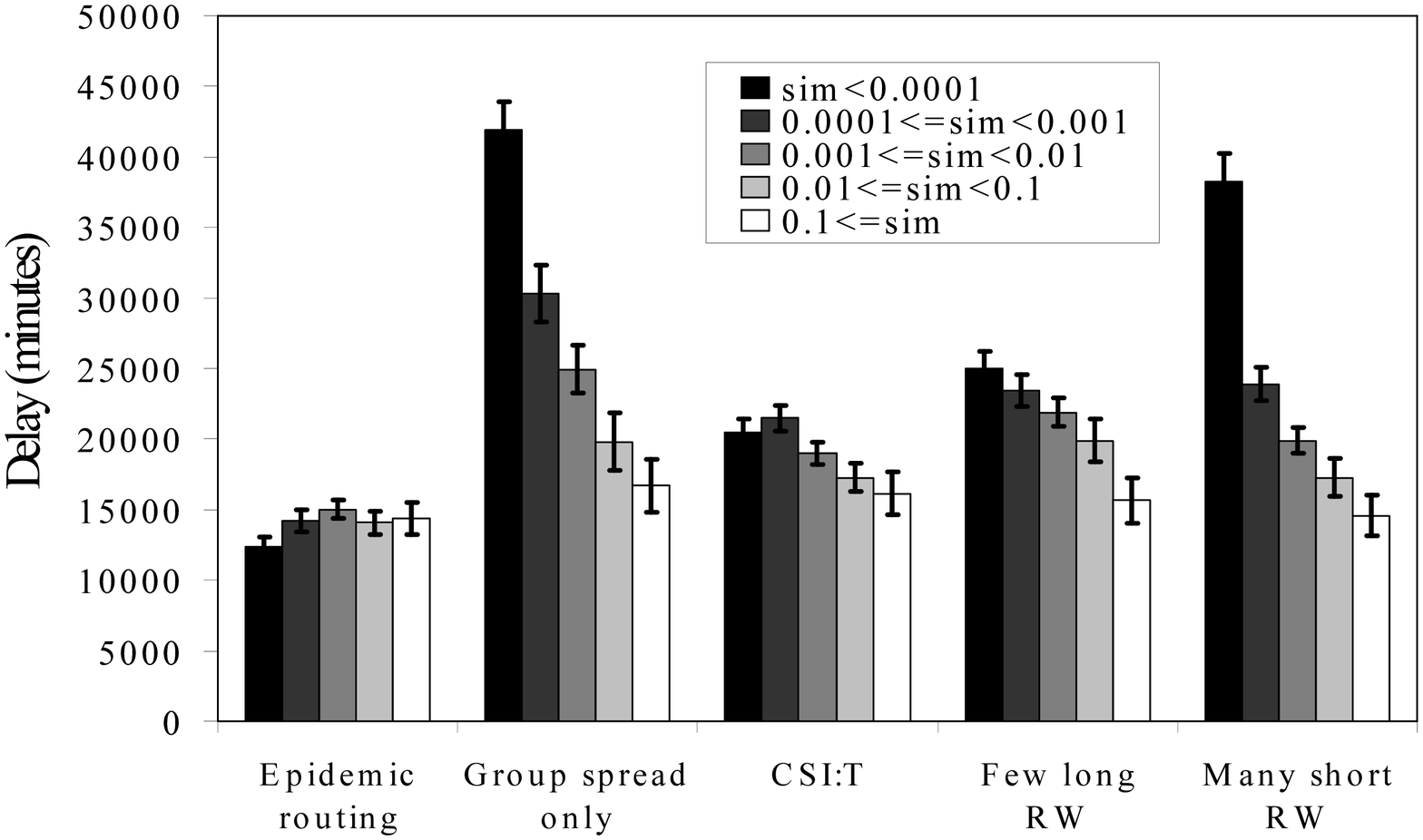} 

\footnotesize{(b) Average delay.}

\caption{Split performance metrics by the similarity between the sender and the target profile (USC).}

\label{MPC_split_stats}
\end{figure}

We take a closer look at the performance metrics by splitting the simulation cases into categories, depending on the original similarity metric between the sender's behavioral profile and the {\it TP}, $Sim(BP(S), TP)$. By the split statistics shown in Fig. \ref{MPC_split_stats}, we see why the {\it gradient ascend phase} is needed to improve the success rate and reduce the delay. When we use only the {\it group spread phase}, and the sender is dissimilar from the {\it TP}, it takes a longer time before any encounter event happens directly between the sender and anyone in the neighborhood of the {\it TP}, if it happens at all -- hence the delay is longer, and the success rate is lower.

\begin{figure}

\centering

\includegraphics[width=3.2in]{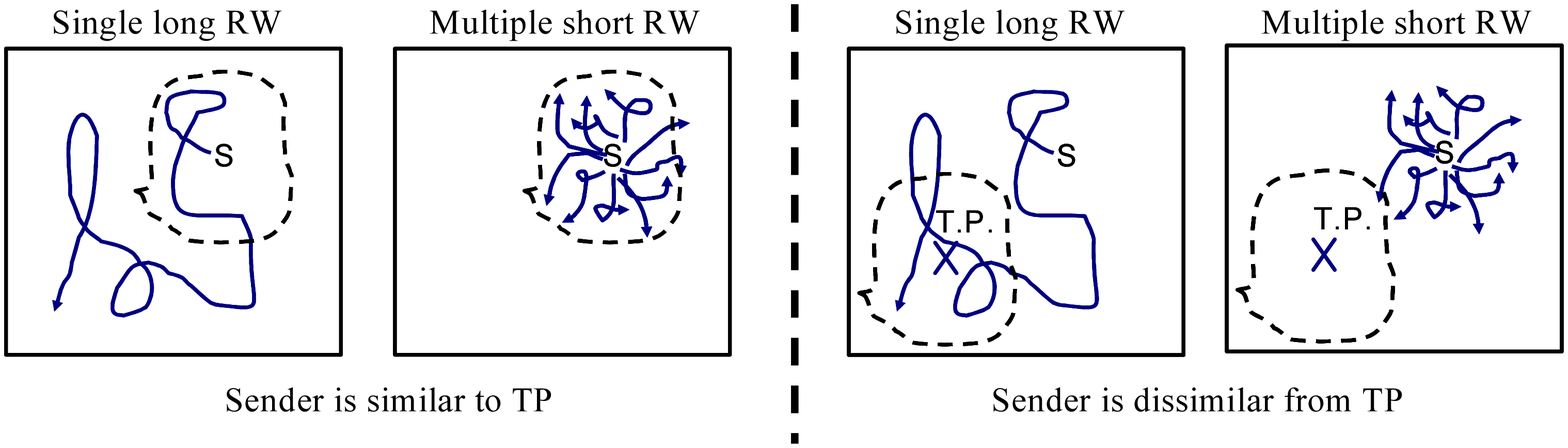} 

\caption{Illustrations for the comparison between one long random walk and many short random walks.}
\label{RW_illustration}
\end{figure}

Comparing the differences between two versions of random walks, few long threads and many short threads, reveals an interesting difference. The concept that leads to the difference is illustrated in Fig. \ref{RW_illustration}. Many short threads are better if the sender is close to the {\it TP}, in terms of both delivery ratio and delay, as the sender generates a lot of threads to ``occupy'' the neighborhood -- since the threads are short, and similar users encounter more frequently, they are likely to stay in the neighborhood. Contrarily, if the sender is far away from the {\it TP}, long random walk threads provide a legitimate chance of moving close to the {\it TP}, while short threads provide less hope.

\subsection{CSI:Dissemination Mode}

\subsubsection{Simulation Setup}

In the scenario of {\it CSI:D mode}, the target
profile specified by the sender cannot help to determine to where
the message should be sent in the behavioral space. Hence, the
strategy seeks to keep one copy in every neighborhood in the
behavioral space. In our evaluation, we start from $1000$ randomly
selected users as the senders. Since the target profile of the intended receivers 
can be orthogonal to the behavioral profile, we create the scenario
for evaluation by randomly selecting $500$ nodes as the intended
receivers for each sender, and consider the average
performances. We vary the two thresholds, $th_{fwd}$ and $th_{nbr}$ in
our {\it CSI:D mode} scheme proposed in \ref{sec:MIPC},
to adjust the aggressiveness of the forwarding scheme. Setting low
values for both thresholds leads to less aggressive operations and
inferior performances. At the same time is also leads to lower overheads, as the messages are copied
to fewer message holders, and the existence of a message holder
prevents nodes in a larger neighborhood from becoming another message
holder.

We compare various parameter settings of our {\it CSI:D} mode
with two baseline protocols, the
{\it epidemic routing} and the {\it random walk}. The epidemic
routing works the same way as before, serving as the baseline for
comparison. In the random walks, the visited nodes along the walks
become message holders and they will later disseminate the messages
further when encountering with the intended receivers. The {\it optimal}
protocol again assumes global view of the network and the knowledge of
the future. Every node in the network knows who the intended receivers are, and
sends the messages to other nodes only if they lead to the fastest delivery
to the message to one of the receivers.

The performance metrics we consider are {\it delivery ratio},
{\it average delay}, {\it transmission overhead}, and, in
addition, {\it storage overhead}. Here the {\it transmission overhead}
refers to the total number of transmissions to reach the message holders and
the intended receivers. The {\it storage overhead} is the number
of eventual message holders that remains in the network after our
scheme is stabilized (recall that some message holders may decide to
cease performing the task if another message holder is found with
similar behavioral pattern in {\it CSI:D}). This is the overall amount of storage
space invested by the nodes collectively to deliver the message\footnote{Typically, only about
a couple dozens of message holders drop the message in the simulation
cases. Even if we have accounted for the temporarily invested storage,
it adds less than $1\%$ additional storage overhead.}. 
In the {\it epidemic routing} and the {\it optimal} protocol, 
all nodes that receive the message hold on to the message for
future transmissions (there is no distinction between the message 
holder and a regular node), hence the transmission overhead and 
the storage overhead are the same.

\subsubsection{Simulation Results}

\begin{figure}
\centering

\includegraphics[width=2.8in]{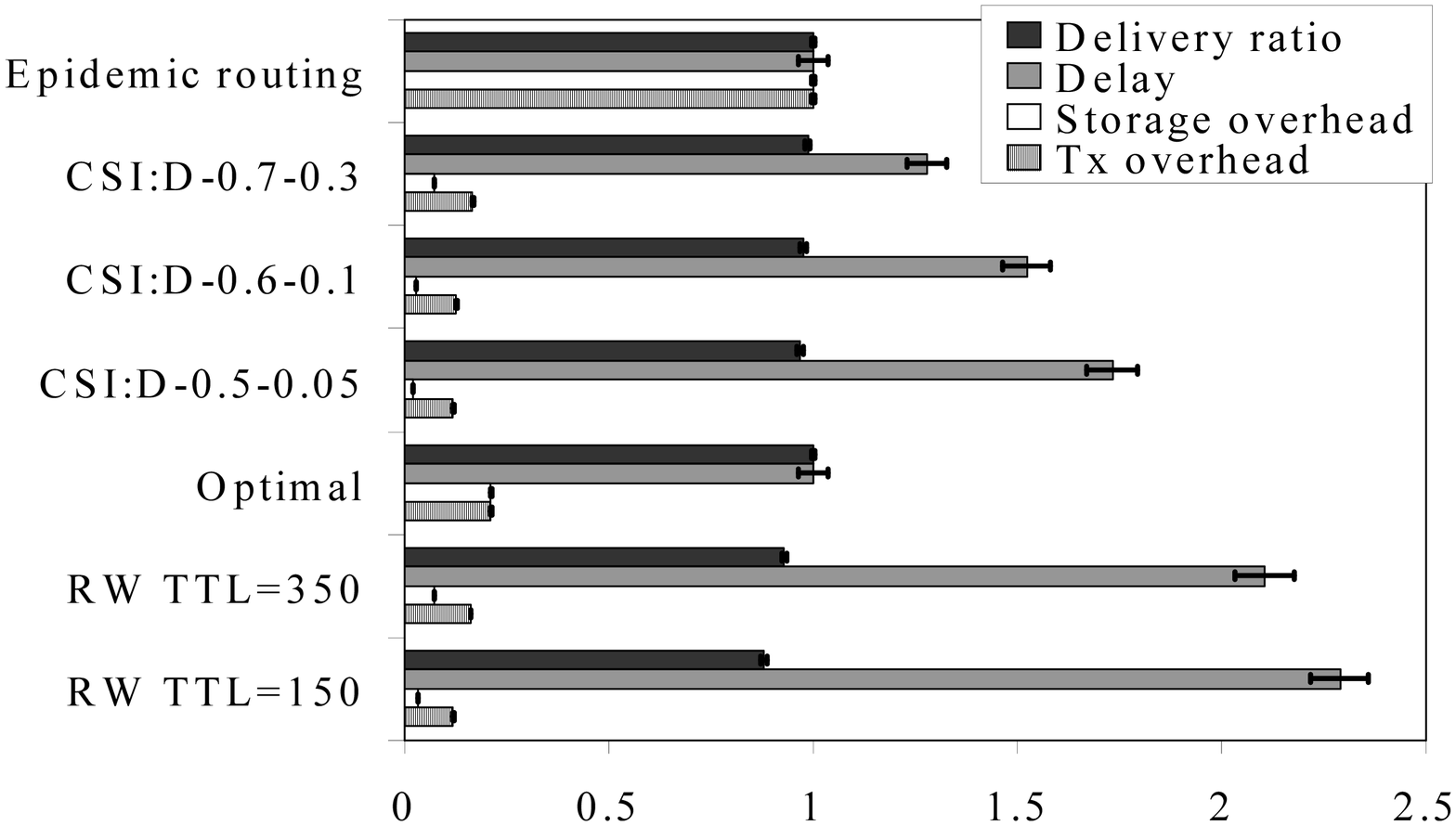} 

\footnotesize{(a) USC.}

\includegraphics[width=2.8in]{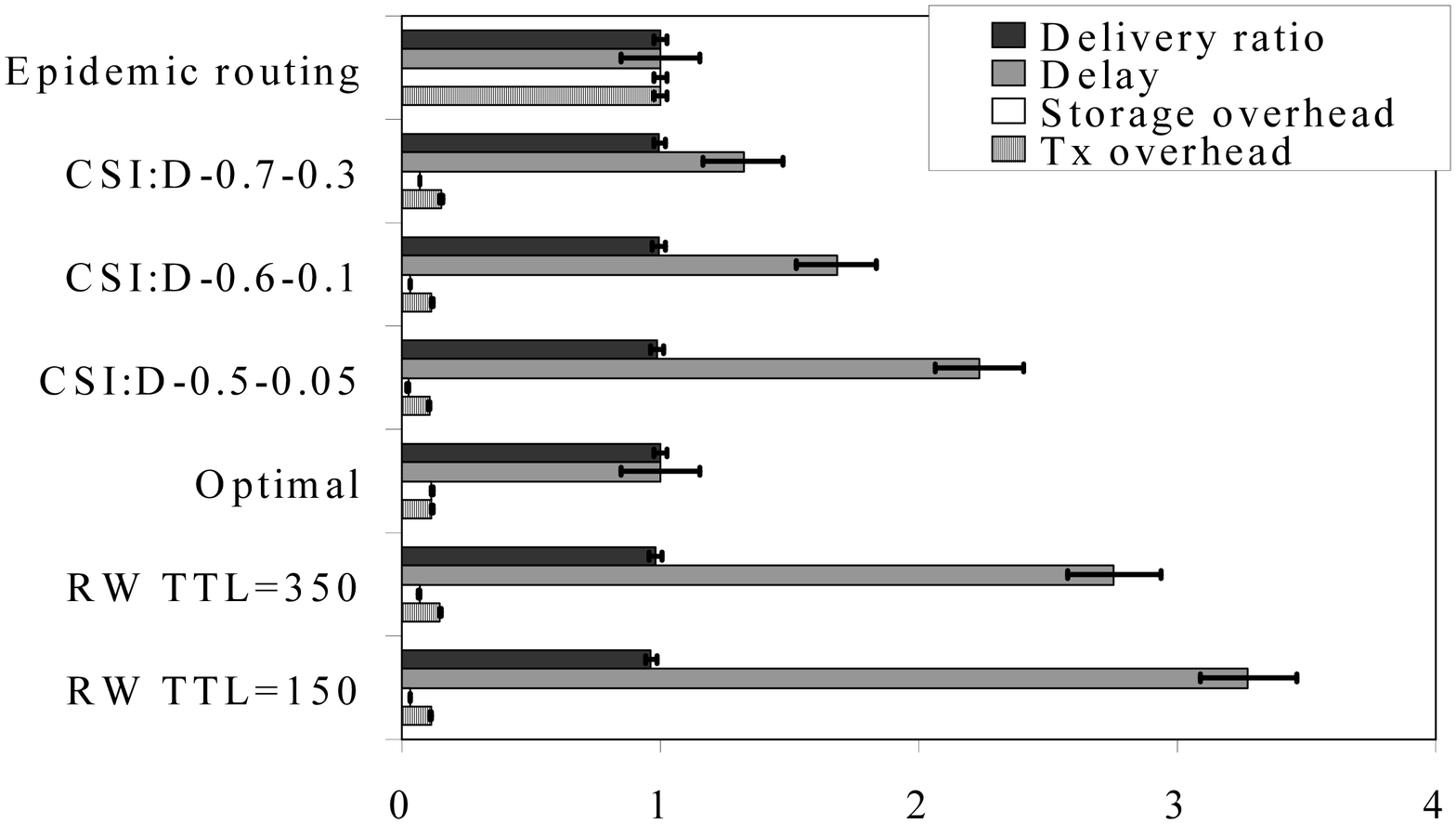} 

\footnotesize{(b) Dartmouth.}

\caption{Performance comparison of CSI:D to other protocols.}

\label{MIPC_performance}
\end{figure}

In Fig. \ref{MIPC_performance} we show the average result of the
$1000$ simulation cases with the $95\%$ confidence interval. We use
the legend CSI:D-$\mbox{th}_{\mbox{fwd}}$-$\mbox{th}_{\mbox{nbr}}$ for our {\it CSI:D}
scheme. Comparing with the {\it epidemic routing}, our protocol saves
a lot of transmission and storage overhead. It is possible to use only
about $7.2\%$ strategically chosen nodes as the message holder and
reach the intended receivers with little extra delay (about $32\%$ more),
when $th_{fwd}=0.3$ and $th_{nbr}=0.7$. Notice that the
storage overhead of the {\it CSI:D} scheme is even lower than the 
{\it optimal} protocol (less than $60\%$) with the objective of minimizing the delay.
If one desires further reduction in the overhead, setting lower threshold values provide a
way to trade performance for overhead, e.g., setting $th_{fwd}=0.1$ and $th_{nbr}=0.6$
cuts the storage overhead to about $3\%$ of the {\it epidemic routing}.
The delay of the {\it CSI:D} is not much more than the {\it epidemic routing} or
the {\it optimal}, at around $27\%$ to $32\%$ more when $th_{fwd}=0.3$ and $th_{nbr}=0.7$.

For the {\it random walks}, we have configured the $TTL$ values for them
to have similar overhead with the {\it CSI:D} (i.e., compare RW TTL=350
with CSI:D-0.7-0.3 and RW TTL=150
with CSI:D-0.6-0.1). We notice that although the delivery rate 
of the {\it random walk} is also pretty good  
($1.5\%$ to $10\%$ inferior to the
corresponding {\it CSI:D}), thanks to the 
non-zero encounter probability between dissimilar nodes, its
delay is much longer than the corresponding {\it CSI:D}
(between $50\%$ to $108\%$ more). This is because the {\it random walk} does 
not leverage the implicit structure of the human network to select
the message holders wisely, as the {\it CSI:D} does.
The {\it random walk} leaves copies within the same neighborhood of the original sender with higher probability, as similar nodes are more likely to encounter (i.e., the {\it random walk} will not ``leave the neighborhood'' in a small number of hops). Hence, there exists significant overlap between the nodes encountered by the selected message holders, and the other nodes that are dissimilar to these holders have to wait for a long time before some ``random'' encounter events occur to receive the message, resulting in the longer delay.


\section{Discussions} \label{sec:disc}

\subsection{Additional Overhead} \label{sec:overhead}

In addition to the message transmission and storage, in our proposed
CSI schemes, due to the need for exchanging and
maintaining the behavioral profiles, there are some additional
overhead. We discuss them in details in this section.

\noindent \textbf{\underline{Overhead for exchanging the behavioral
profiles}} We identify some additional components to the actual
message transmissions when the encounter events between mobile nodes
are leveraged for message dissemination. Some of the components are
common to {\it any} message dissemination schemes, and the others are
unique to our CSI schemes.

\begin{itemize}

\item The common overhead for all the DTN message dissemination schemes
considered include
the beacon signals for nodes to discover each other when they
encounter, and the exchange of a list of ``messages I have seen'' to
avoid a given node receiving duplicated messages from different
nodes. This type of overhead is a function of the encounter patterns itself
and is independent of the actual protocol used. We ignore these common factors in our analysis.

\item Exchanging the behavioral profiles for the evaluation of mutual
similarity is an additional component that exists only in our
behavior-aware protocol. These profiles are a handful of vectors
associated with its weights. For most of the users, empirically, 
five to seven eigen-behavior vectors capture more than $90\%$
of the power in their {\it association matrices}~\cite{MOBI07}.
This is a small constant overhead we pay for
each encounter when one of the nodes has some message to send. 
If the message size is much larger than the overhead, which is
usually the case as messages are transferred in a bigger
unit (i.e., a ``bundle'') in DTNs, it is
worthwhile to pay this overhead to gain the reduction of transmission
counts as we see in section \ref{sec:exp}. Furthermore, with CSI, if there
is no message to send, there is no need to exchange the behavioral
profile. Thus, comparing with the protocols that require proactive, persistent 
exchanges of control messages when nodes encounter (e.g., ProPHET~\cite{PRoPHET}
requires the exchange of encounter probability vectors),
qualitatively, the CSI schemes have lower overhead, especially when
the volume of traffic is low in the network.

\item The actual message size has to be augmented with the {\it TP}
as well. This is a constant
overhead, and it can be reduced if the target vector is ``sparse''
(e.g., if the {\it TP} considers only the visits to the gym exclusively,
there is only one $1$ in the vector. Instead of adding a vector
$(0,...,0, 1, 0, ....)$ in the header, the vector can be encoded 
(i.e., by specifying (gym, 1)) to save space.).

\item In the CSI:D mode, the message holders
have to exchange the list of behavioral profiles of known
holders. This happens only between a small subset (less than $8\%$) of
the nodes, and the exchange is necessary only when there is a
difference in the lists. To further alleviate this, the two nodes can
compare their known holder lists using a hash value, and
exchange only the difference.
\end{itemize}

\noindent \textbf{\underline{Overhead for maintaining the behavioral
profiles}} In order to maintain the behavioral profile, the nodes have
to keep track of its visiting time to various locations. Note this does not
require a node be aware of all possible locations in the environment -- 
it has to keep track of only the ones it has been to. 
When two nodes exchange the behavioral
profiles, each entry in the behavioral profile contains only a 
subset of locations with annotations for these locations (e.g., Node $A$
specifies (library, gym) = (0.8, 0.2) while node $B$ specifies
(library, computer lab) = (0.4, 0.6)).
The nodes will take a union of the location sets
when comparing their similarities (e.g., in the previous example,
when node $A$ sends the behavioral profile to $B$, $B$ will
convert the profiles to $BP(A)$: (library, gym, {\it computer lab}) = (0.8, 0.2, 0)
and $BP(B)$: (library, {\it gym}, computer lab) = (0.4, 0, 0.6)
before comparing). The required storage on each node is minimal,
as we show about three to five days of summarized {\it mobility
preference} is sufficient to establish a stable behavioral profile for the user
in section~\ref{sec:mobipattern}.

In addition, if the beacon signals from locations are not available,
it is possible to use the mutual encounter vectors as the behavioral
descriptors for the nodes -- nodes who move similarly should have
similar encounter sets. In this sense, we could replace the representation
to be totally independent of the infrastructure.

\subsection{Privacy Issues}

While the behavior-aware message dissemination schemes achieve good
performance with significant overhead reduction, it also raises
user privacy concerns. In some cases, individuals may not want
to reveal their own behavior. We discuss privacy-preserving options
with our CSI scheme below.

First we emphasize that the original design of CSI presented in section~\ref{sec:proto}
inherently possesses a privacy-preserving feature: we only use a small
subset of user behavior (specifically, the mobility preference)
in the behavioral profile, and with the singular value decomposition, 
we reveal only the summarized trend, not detailed location visiting events
for the user. In addition, the behavioral profiles are
exchanged only between nodes, not stored in any public directory, and
it limits only to when a given node is involved in message
dissemination. 

We can further reduce the behavioral profile exchanges in the CSI
scheme, and hence help to preserve privacy as follows.
For the CSI:T mode, when nodes encounter, instead of
exchanging their behavioral profile, the node with a message to send
would first send to the other node the {\it TP} of the message and its similarity score to the {\it TP}.
The other node silently calculates its similarity to the {\it TP} and decides whether
to request for the actual message. This completely removes the need
for behavioral profile exchanges in CSI:T mode.

For the CSI:D mode, when a message holder looks for potential new holders,
instead of asking other nodes to send the behavioral profile, the message
holder sends the list of known holder's behavioral profiles to the other node.
Since this list contains only the {\it behavioral profiles} of the known holders,
not their {\it identities}, dissemination of such lists in the network does not pose
a threat to the privacy of the message holders. Furthermore, when there are 
multiple holders in the list, the other node is not able to tell which behavioral
profile corresponds to the holder who sends out the list. If the other node decides
to become a message holder, its behavioral profile has to be added to the 
list of known holders. Instead of immediately sending the behavioral profile of the new
holder to the old holder, which poses an opportunity for the old holder
to link the identity and the behavioral profile of the new holder, the new holder
only adds its behavioral profile to its own known holder list, and delays the dissemination
for a later holder profile list exchange.

Finally, as a last resort, privacy-minded individuals can always opt-out of the service, and we
expect this would not impact the performance severely, as it has been
shown that the encounter pattern between nodes in mobile networks is
rich enough to sustain up to $40\%$ of nodes opting out before observing
a performance degradation~\cite{group-study}.

\section{Conclusion and Future Work} \label{sec:conc}

In this paper, we propose a
paradigm to represent, summarize and manipulate behavioral profiles and use such
profiles as targets for the communication. We have presented
a novel service of message
dissemination in infrastructure-less mobile human networks based
on the behavioral profiles of the users. 
The CSI schemes meet the design goals outlined in section~\ref{sec:design_goal}
with respect to efficiency, flexibility and privacy preserving properties.
The CSI schemes perform
closely to the delay-optimal protocols
(with $94\%$ or more success rate, less than $83\%$ of overhead,
and the delay is inferior by $40\%$ or less). In
addition, we also observe that human behavior as observed in the large
scale empirical traces is quite robust and only a few days' worth of
data is adequate to summarize and leverage for message dissemination,
which is quite surprising.

We are working toward an implementation of the
CSI schemes based on mobile devices and consider a real-world evaluation.
One key issue is to adapt our algorithm in a more privacy-preserving fashion which is also
resistant to spam (e.g., include a reputation system). We are also considering different applications of
behavioral profiles, including targeted advertising via our CSI schemes.

\end{document}